# Hypersharp Neutrino Lines

R. S. Raghavan

*Department of Physics & Institute of Particle, Nuclear and Astronomical Sciences,*
*Virginia Polytechnic Institute and State University, Blacksburg VA 24060*

Neutrino ($\nu_e$) lines (and γ-rays) from *very long lived* nuclei in simple crystals such as metals have hypersharp *natural* width Γ, motionally narrowed by lattice vibrations in analogy to recoilless emission. A generalized hypersharp line fraction including the recoilless part can be derived in a frequency modulation approach. The $\tilde{\nu}_e$ lines of natural width in $^3H \leftrightarrow ^3He$ 2-body β-decay can then be resonantly captured with geometrical cross section. The extreme sharpness $\Delta E/E \sim 10^{-29}$ of the tritium $\tilde{\nu}_e$ line can probe the Planck length $\mathcal{L}$ via its limits on the widths of states, $\Delta E/E(\mathcal{L}) = \mathcal{L}(\mathcal{L}/R)^\beta = 10^{-20}$ (β ~1) to $10^{-40}$ (β= $\mathcal{L}$/R(fm)). Stringent limits can be set on β, thus, on models of quantum gravity.
(May 25, 2008)

Sharp nuclear γ-ray lines have been observed via the Mössbauer effect (ME), the narrowest line attempted so far being that from the τ = 9.3μs isomer of $^{67}Zn$ [1] with $\Delta E/E = \Gamma/E_\gamma < 10^{-15}$ (the natural width $\Gamma = \hbar/\tau$). The ultimate line width Γ is realized only if the nuclear state is free of perturbations that disperse the line energy beyond Γ. In practice, the state is subject to dynamic and static perturbations. Spin relaxation of nearby nuclear and electronic moments is a prime example. With broadening due to a fluctuation time T, one expects to observe a width Ω ($=\hbar/T$) > Γ. This dilutes the spectral density at the resonance energy, thus also the cross section σ for resonant absorption as:

$$\sigma = \sigma(\Omega) \sim \sigma(\Gamma)/\Omega \ll \sigma(\Gamma) \qquad (1).$$

For the maximum resonant spectral density 1/Γ, σ(Γ) rises to the full geometrical value [2] $2\pi^2 \lambdabar^2$. Generally, T is $10^{-3}$ to $10^{-10}$ s, and τ of ME levels is in the same range. Thus line widths Ω ≥10 Γ that reduce σ(Γ) by factors of 10 are common.

In the quest for γ-rays sharper than the $^{67}Zn$ line, attempts been made to observe the ME in isomers in $^{107,109}Ag$ with τ~40 s thus, a natural width ~$1.5 \times 10^{-17}$ eV. Under the experimental conditions, severe line broadening is expected from (1), largely via dipolar field fluctuation at a rate 1/T ~ 10 kHz. Thus, Ω ≥ $10^{-12}$ eV >>$10^{-17}$ eV. By (1), σ is reduced at least by ~$10^5$, suggesting that hypersharp γ-ray lines in $^{107,109}Ag$ may be out of reach.

Yet, since 1979, several experiments (all difficult), claim to have observed this resonance[3] with σ consistent with a line broadening *only by a factor <10* instead of by ~$10^5$. The results remain controversial. Neither is a satisfactory theoretical basis yet available for their possible validity.

Regardless, the possibility of hypersharp lines from long lived states is key to diverse topics with potentially major scientific impact. Examples are γ-ray superfluorescence [4] and resonant neutrino capture for revolutionary table-top scale neutrino experiments[5,6]. Neutrino emission, by nature, is of course, very long lived (τ~ days to years).

In this Letter I examine the basis of line-broadening *for very long-lived states*. I conclude that γ-rays and $\nu_e$ from such states can be emitted without broadening from simple lattices in analogy to the ME itself. For fluctuations with $\hbar\Omega \gg \Gamma$, a condition always satisfied for long lived states (but not in ME lines observed so far, all being short lived), γ-rays and $\tilde{\nu}_e$ lines from 2-body β-decays e.g. from tritium[5,6], will be emitted with the *natural* line width. Then, the tritium $\nu_e$ width $\Delta E/E \sim 10^{-29}$, is some $10^{15}$ times sharper than ME lines to date. In this case, σ(resonance) nears the geometrical limit σ~$10^{-18}$ cm$^2$, some $10^{25}$ times larger than σ of usual $\nu_e$ reactions. Both aspects, if realized, will revolutionize ν research and open deep questions of nature to experimental test.

In a simple crystal, e.g. a metal, with atoms fixed at lattice sites, atomic motion is controlled by lattice vibrations. The nuclei sense a dipolar field $\mu_1 \cdot \mu_2 /r^3$ ~$3 \times 10^{-12}$eV(r is the interatomic distance). Even in a rigid lattice this (inhomogeneous) field is *not static* because r and thus the field intrinsically fluctuate via lattice vibrations (recall that <x> ~1A° ~ r from recoilless emission) The key idea here is that we expect these THz harmonic vibrations to average out the dipolar field so that the line energy is sharp. Stochastic motion--random jumps common in liquids and molecular solids--cannot be averaged away in general. They could coexist with vibrational motions but decoupled from each other due to the vastly different (THz vs KHz) time scales. We assume here that at least part of the dipolar fluctuation is vibrational, thus *harmonic*.

Salkola and Stenholm[7] considered the effect of harmonic fluctuations on ME line shapes (including the role of the nuclear life time) as a frequency modulation (FM) of the emitted line using a modulation model $\Omega_o \cos\Omega t$. This simple model with one frequency is a template of the effect of the complex lattice vibration spectrum but contains the essential features of the problem. We adopt this model with $\Omega_o$ the (dipolar) field splitting and $\Omega$, its fluctuation rate. The line shape is then[7]:

$$A \propto \frac{1}{\Gamma} \Sigma_{k=-\infty}^{k=+\infty} J_k^2(\eta) \frac{1}{[(\delta/\Gamma) - k\xi]^2 + 1} \quad (2),$$

where $J_k(x)$ are Bessel functions, $\eta = \Omega_o/\Omega$, $\xi = \Omega/\Gamma$ and $\delta$ is the external detuning for scanning the line shape. For the central line, $\delta = 0$. The signal consists of a central line and an infinite series of sidebands of index $\pm k$, all with the natural width.

The broadening of the central line is due to the overlap of the first sidebands with $k = \pm 1$ with the $k=0$ term in (2). The first sidebands occur at the energy $\delta$ tuned to make the square bracket in the denominator in (2) zero. The first side band thus occurs shifted from the central line by $\delta/\Gamma = \xi$ *linewidths*. Thus, larger the $\xi = \Omega/\Gamma$, less the overlap, even though the absolute value of $\delta$ eV of the $k=1$ sideband position is set only by $\Omega$. A sideband offset $\xi$ improves by increasing $\Omega$ (e.g., by higher ambient temperature etc). However, $\xi$ increases *naturally* also because *$\Gamma$ decreases* in long lived states. A narrow central line of natural line width, well resolved from the side bands is thus achieved naturally and *necessarily* in the case of long lifetimes. As shown long ago directly as a way avoiding line broadening in ME[8], the same effect results for an *external* oscillating field. In contrast to long lived states, short lived states (small $\xi$) emit a wide central line as well as sidebands. They are thus poorly resolved and lead to line broadening.

For example, in the case of the Ag isomers, $\Gamma \sim 10^{-17}$ eV and $\hbar \Omega \sim 10^{-12}$ eV (10 kHz usually cited as "relaxation" time), the $k = \pm 1$ sidebands occur $10^5$ natural line widths away, leaving a highly isolated central line of natural width regardless of the sideband intensities. In contrast, for the same $\Omega$, in the $\tau = 10$ μs state in Zn, the natural widths for all k is $10^7$ times broader, $\xi = 0.01$ thus the k > 0 sidebands on either side overlap almost completely with the central line and broaden it towards rule (1).

Observed line widths (and resonance cross sections) are thus *different in long lived states* from those in short lived states in the same matrix and subject to the same perturbations. On the other hand, the *intensities* of all the terms in (2) given by the $J_k(\Omega_o/\Omega)$ are entirely independent of the lifetime. They are determined only by the fluctuation parameters. For example, for $\eta = 1$ ~58% of the intensity is in the central line and ~38% in the $k = \pm 1$ sidebands for *all nuclear states, short or long lived*. Lattice vibrational theory in detail may change the intensities in the central and sideband lines, not the line widths. As mentioned before, in the vibrational model, $\Omega$ is large (THz range), thus $\eta << 1$ and $J_0 \approx 1$.

The relevance of the FM perspective of γ-ray line shapes was shown long ago by Shapiro[9] for the ME itself. He treated the fluctuating Doppler shift of the emitted line due to nuclear motion in a vibrating lattice. This affects the phase of the wave of the transition as $\varphi(t) = \omega_o t + x(t/\lambdabar)$. The instantaneous displacement $x(t)$ can be expanded in a series of the lattice vibration frequencies $\Omega_m$. Then the wave field E is:

$$E = e^{i\varphi t} = exp\, i[\omega_o t + \Sigma_m \frac{x_m}{\lambdabar} sin(\Omega_m t)]$$

$$= \prod_m \Sigma_n J_n(\frac{x_m}{\lambdabar}) exp[i(\omega_o + n\Omega_m)t]$$

The fraction of the unshifted line (Debye-Waller) factor) with n = 0 is given by[9]

$$f = E^2 = \prod_m J_o^2(x_m/\lambdabar)$$

Since $x_m/\lambdabar = x_m\omega_o/c < 1$, and $J_o(y) \approx 1 - y/$

$$f = 1 - \Sigma(x_m/\lambdabar)^2 = exp-[<x>^2 \lambdabar^2] \quad (3)$$

where $<x>^2 = \frac{1}{2} \Sigma(x_m)^2$. The central unshifted line (the ME line) is accompanied by an infinite number of sidebands at $\omega_o \pm n\Omega_m$ each of intensity $J_n$. The ME line width depends, as discussed above, on the overlap from the first sidebands. As noticed *already in 1962*[10], the condition for side band resolution is a *lifetime longer than $1/\Omega_m$*, explicitly anticipating the basic conclusion above. For a typical ME case, $\tau >> (1/\Omega_m) \sim 10^{-13}$ s, thus the condition for sideband resolution is automatically satisfied in every ME resonance observed so far.

Other causes of line broadening arise differently. The line energy may be shifted by fixed amounts from the line energy due to many reasons. The equilibrium values of many of these shifts Δ e.g., chemical shifts $\Delta_c$ depend on the nuclear volume $(1/r^3)$ that encloses the charge which could differ in the excited and ground state (isomer shifts). The shift itself is relatively fixed but only up to fluctuations that broaden $\Delta_c$ which ultimately broadens the line itself. The effect may appear small but it is relevant in the hypersharp context. Since $\Delta_c$ depends on the position coordinate r, it will fluctuate (as before in the case of dipolar fields) due to lattice vibrations. The FM picture with vibration frequencies $\Omega_m$ can thus be applied here also. The broadening is determined again by $\xi = \Omega_m/\Gamma$ which, for long lived states, can be very large, thus restoring the hypersharp line. The intensity of the central line is $J_o(\delta\Delta_c/\Omega_m)$, practically undiluted since e.g., the usual isomer shift in known cases is itself $\sim 10^{-6}$ eV. Thus the deviation $\delta\Delta_c$ is at worst, of the same. Since $\hbar\Omega_m \sim 10^{-2}$ eV in the vibrating lattice, the hypersharp line fraction $J_o(\delta\Delta_c/\Omega_m) \approx 1$.

Another source of broadening particularly for hypersharp *neutrino* lines is the gravitational red shift that oscillates with the vibrating atom. The red shift is $\Delta_g/E = gz/c^2 = 10^{-18}$ /cm where z is the vertical displacement. The mean displacement of the nucleus may be inferred from the Debye-Waller factor $f = \exp -<x>^2/\lambdabar^2$ derived above. For a typical low energy ($\sim 20$ keV) line $<z> \sim 10^{-8}$ cm. Thus, the mean fractional red shift $\Delta_g/E \sim 10^{-26}$. Normally such a tiny effect is entirely irrelevant. However for hypersharp neutrino lines e.g, in the two-body decay of tritium, $\Gamma/E \sim 5\times 10^{-29}$ or an internal red shift of $\sim 500$ line widths. This shift will also be motionally narrowed via the lattice vibrations in close analogy to the recoilless effect. The relevant parameters $\xi = \Omega_m/\Gamma >> 1$ and $J_o^2(\Delta_g/\Omega_m) \approx 1$ in this case assure a hypersharp neutrino line. External red shifts occur however, due to the macroscopic source/absorber geometry, needing further remedy (see below) via experimental design.

A common theme emerges from these discussions for all (harmonic) fluctuations that affect hypersharp lines defined arbitrarily as those with lifetimes $\tau > 1$ s, thus $\Gamma < 10^{-15}$ eV. The rate $\Omega$ of energy fluctuations of any known process in a crystal is always many orders of magnitude larger than $\Gamma$. Thus $\xi = \hbar\Omega/\Gamma$ is guaranteed to be >>1, thus also hypersharp lines of natural line width. Secondly the intensity of the final line is a product of probabilities of each source K of fluctuation $J_o(\Delta_K/\Omega)$ where $\Delta_K$ is the energy width of the fluctuation and $\Omega_K$ is its rate. Thus we can define a generalized hypersharp line fraction $\mathcal{H}$:

$$\mathcal{H} = J_o^2(<x>/\lambdabar) \; \Pi_K \; J_o^2(\Delta_K/\hbar\Omega_K) \qquad (4)$$

where K runs over the different types of fluctuations with width $\Delta_K$ and rate $\Omega_K$. The usual recoilless fraction f, the first term, is now only one of many that determine the hypersharp line intensity. Eq. (4) can be used to design hypersharp lines in practice by optimizing parameters individually for each source of fluctuation.

In general, motional averaging can occur if the fluctuations are *symmetrical* i.e. energy is added and subtracted to the line about zero equally probably and the width of the swing covers the perturbation width. Thus e.g. while the first order Doppler effect ($\propto v/c$) can be averaged to result in the ME line, the second order Doppler effect $\propto v^2/c^2$ cannot. Indeed, it produces a well-known energy *shift* (not broadening). That can be canceled if the source and absorber are held at the same temperature.

Additional line broadening arises from at least two external sources-- macroscopic geometries and the earth's magnetic field. Both these effects result in symmetric fluctuations, thus they can be averaged by suitable design. A gravitational red shift $10^{-18}$ eV/cm $>>\Gamma$ occurs in macroscopic sources and distinct absorbers. Each point in the source has a conjugate point in the absorber at the same gravitational potential. A symmetric vertical motion allows each source point to scan all the conjugate points in a time short compared to the lifetime. For cm size sources /absorbers with a maximum energy spread due to the red shift of $\Omega_o = 10^{-18}$ eV. Vertical motion at a rate $\Omega = (10^{-18} \text{eV}/\hbar) \sim 10^{-3}$ cm/s may be adequate. Similar considerations apply to the earth's field which results in a splitting $\sim 10^{-14}$ eV/$3\mu_N$/mGauss which can be canceled statically to a high degree. If this is done so there are no variation in cancellation over the source –absorber geometry then it is fixed shift and canceled in the forward to reverse transitions. A distribution of small residual fields can then be averaged by swinging the canceling field in direction and intensity about the zero field.

Static interactions are not time dependent. Quadrupole coupling from static *distributed* field-gradients from random crystal defects in the vicinity of the probe (even in a cubic lattice) are the most problematic. This produces irremediable inhomogeneous broadening not subject to motional narrowing. Extreme care in crystal preparation is thus essential but even this may not be sufficient in the hypersharp context. The only guarantee is to eliminate the possibility of this broadening altogether via zero quadrupole moments of *both* excited and ground states. This is the case for spin ½ of the nuclear level. However, such spins do not favor long γ-decay lifetimes that result from large differences of excited and ground state spins; at least one level will most likely have a non-zero quadrupole moment. The sharpness of γ-ray lines may thus be limited to shorter lived states (broader width) and by the prevailing inhomogeneous broadening.

Neutrino lines can connect spin 1/2→1/2 states. Thus the fearsome inhomogeneous quadrupole broadening of such $\tilde{\nu}_e$ lines is *absent*. A case in point is the $\tilde{\nu}_e$ line in the 2-body decay $^3H(1/2)^+ \leftrightarrow {}^3He(1/2)^+ + \tilde{\nu}_e$ ; [(E($\tilde{\nu}_e$)=18.6 keV; τ ($^3$H) ~ 6x10$^8$ s). I have recently proposed[5] an experiment to observe resonant capture of this $\tilde{\nu}_e$. With Γ ~ 10$^{-24}$ eV it is the sharpest of likely hypersharp $\tilde{\nu}_e$ lines. Technical details can be found in the companion paper[6]. Key design concepts are mentioned here.

With motional narrowing of time dependent perturbations as discussed above, the major design problem in the resonant $\tilde{\nu}_e$ capture experiment is the elimination of energy shifts from static sources via *site uniqueness and site identity* of the parent $^3$H and daughter $^3$He in the source and vice versa in the absorber. This is a difficult problem since in almost all materials the daughter He diffuses rapidly and forms bubbles, an ambience very different from $^3$H lattice sites. The pivotal discovery in ref. 5 is a unique material that assures site identity and uniqueness --Nb$^3$H below 200K. Only in this case does $^3$H and $^3$He occupy the same unique interstitial lattice site in identical sources and absorbers *All* fixed energy shifts Δ in the emission are then *exactly compensated for* by opposite shifts in the reverse transition. Examples of the fixed shifts that are so compensated are: the second order Doppler effect (SOD), chemical shifts and possibly others[5,6].

The cross section σ(Γ) for this $\tilde{\nu}_e$ resonance of natural width, is simply the geometrical limit σ ~10$^{-18}$ cm$^2$ (including a recoilless fraction f$^2$ = 0.076). (The much smaller value in ref. 5 includes unjustified dilution by "relaxation"-broadening). The dramatically large σ for a $\tilde{\nu}_e$ reaction (in the class of x-rays) implies mg/cm$^2$ thick "black" $\tilde{\nu}_e$ absorbers! This cross section, ~10$^{25}$ larger than that for conventional $\tilde{\nu}_e$ reactions, bodes well for precise table top scale experiments[5,6] for tritium $\tilde{\nu}_e$ oscillations via $\theta_{13}$ and active-sterile mixing[11].

We have surveyed many sources that affect the energy, width and intensity of hypersharp lines. Still, considering the huge jump in precision, other sources cannot at all be ruled out. The reason for a measure of optimism is the huge cushion of σ~10$^{-18}$ cm$^2$ that can withstand many of them. For example, I assume harmonic spin dynamics justified by metal lattices, but a general lattice may include the presence of (uncorrelated) stochastic motion. Even if the harmonic motion due to lattice vibrations is only 1% of the relaxation, it only reduces the signal *rate,* not the spectral width. A x100 smaller signal is entirely acceptable with the huge reserve in σ.

Hypersharp tritium $\tilde{\nu}_e$ lines open new paths for probing deep particle and fundamental physics questions Such developments were seen after the discovery of the ME that offered the precision ΔE/E~10$^{-15}$. This is now dramatically enhanced to ΔE/E~10$^{-29}$, challenging the imagination of new physics perspectives that such a tool could unveil.

For example, Mead[12] suggested that a fundamental length (Planck length) $\mathcal{L}$ in nature would limit the ultimate widths of nuclear states i.e. ΔE/E depends on $\mathcal{L}$. With the definition $\mathcal{L}$= (G ℏ/c$^3$)$^{1/2}$ ~10$^{-33}$ cm, Mead predicts ΔE/E($\mathcal{L}$) =$\mathcal{L}$ ($\mathcal{L}$/R) β ~ 10$^{-20}$ (for β = 1) to ΔE/E($\mathcal{L}$) ~10$^{-40}$ (for β=$\mathcal{L}$/R) (R is the nuclear radius). The form of β (in general ($\mathcal{L}$/R)$^n$), depends on the quantum gravity model.

Detecting this effect needs a hypersensitive probe. The hypersharp tritium $\tilde{\nu}_e$ resonance offers such a tool. Indeed, a width ΔE/E~10$^{-20}$ implies a Planck broadening by 10$^9$ of the tritium $\tilde{\nu}_e$ resonance. A dilution of σ by this large a factor can be easily detected. Thus a measurement of σ is itself the first test of the range of ΔE/E ($\mathcal{L}$). Observation of the $\tilde{\nu}_e$ resonance would already preclude β = 1. Further results from scans of the tritium $\tilde{\nu}_e$ resonance can influence quantum gravity models.

Many known nuclides live longer than implied by the width limit ΔE/E~10$^{-40}$ which calls for reconciliation with the time-energy uncertainty relation.

I wish to thank L.N. Chang, D. Minic M.Pitt W. Potzel and F. Wagner for helpful discussions.